\documentclass[
journal=jacsat, 
manuscript=article]{achemso}

\usepackage[version=3]{mhchem}
\usepackage{latexsym}
\usepackage{amsmath}
\usepackage{amssymb}
\usepackage{graphicx}
\usepackage{caption}
\usepackage{subfigure}
\usepackage{float}
\usepackage{mathrsfs}
\usepackage{color}
\usepackage{epstopdf}

\author{Zhedong Zhang}
\email{zhedongz@uci.edu}
\affiliation{Department of Chemistry, University of California Irvine, Irvine, CA 92697, USA}
\author{Kochise Bennett}
\affiliation{Department of Chemistry, University of California Irvine, Irvine, CA 92697, USA}
\author{Vladimir Chernyak}
\affiliation{Department of Chemistry, Wayne State University, Detroit, MI 48202, USA}
\author{Shaul Mukamel}
\email{smukamel@uci.edu}
\affiliation{Department of Chemistry, University of California Irvine, Irvine, CA 92697, USA}

\date{\today}

\title{Utilizing Microcavities to Suppress Third-order Cascades in Fifth-order Raman Spectra}

\begin{document}


\begin{abstract}
Nonlinear optical signals in the condensed phase are often accompanied by sequences of lower-order processes, known as cascades, which share the same phase matching and power dependence on the incoming fields and are thus hard to distinguish. The suppression of cascading in order to reveal the desired nonlinear signal has been a major challenge in multidimensional Raman spectroscopy, i.e., the $\chi^{(5)}$ signal being masked by cascading signals given by a product of two $\chi^{(3)}$ processes. Since cascading originates from the exchange of a virtual photon between molecules, it can be manipulated by performing the experiment in an optical microcavity. Using a quantum electrodynamical (QED) treatment we demonstrate that the $\chi^{(3)}$ cascading contributions can be greatly suppressed. By optimizing the cavity size and the incoming pulse directions, we show that up to $\sim$99.5\% suppression of the cascading signal is possible. 
\end{abstract}

\begin{figure}
 \centering
   \includegraphics[width=2in,height=2in]{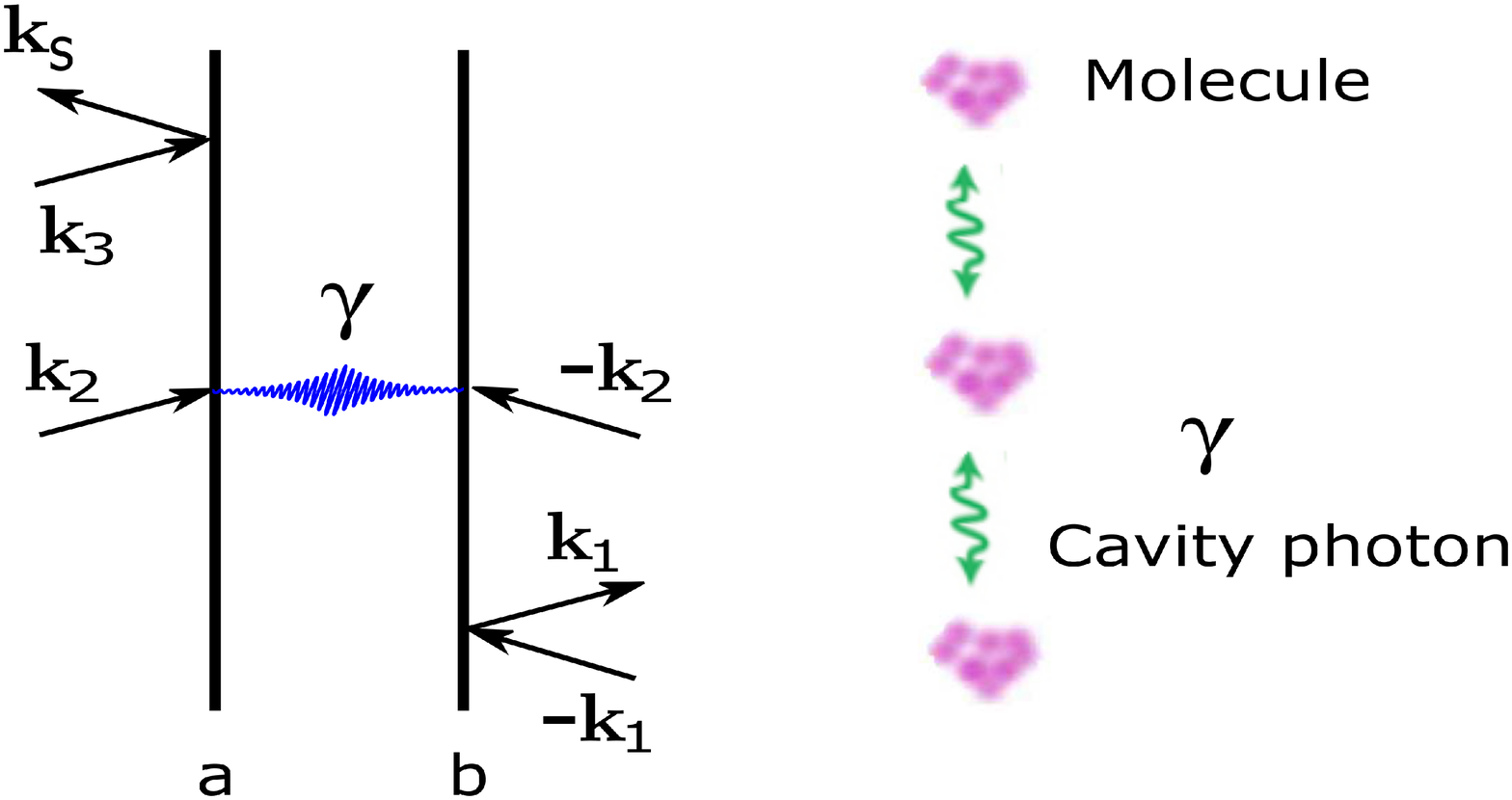}
\caption*{Table of content.}
\label{TOC}
\end{figure}

\maketitle

Multidimensional nonlinear optical spectroscopy provides a wealth of information beyond linear techniques, which can only access the single-excitation spectrum. Multidimensional Raman spectroscopy is an effective tool for studying molecular vibrations and offers a fingerprint by which molecules can be identified.  
However, a many-body effect known as cascading often contaminates Raman spectra in condensed phases and has been the main obstacle in the development of multidimensional Raman spectroscopy \cite{blank1999,mehlenbacher2009,Miller_JCP2008}. Various techniques for separating out these processes have been developed \cite{golonzka2000, blank1999,zhao2011,frostig2015single,gelin2013simple}. Recently a microscopic QED treatment of cascading was developed which connects it to virtual photon exchange between molecules and was applied to various sample geometries \cite{cascadingQED,bennett2014cascading}. A host of other effects owe their origin to the quantum nature of the electric field. These include local-field effects \cite{bennett2014cascading,mukbook,lozovoy2001cascaded,cundiff2002time}, dipole-dipole coupling \cite{thirunamachandran1980intermolecular,salam}, the Lamb shift \cite{scully2009collective}, induced nonlinearities \cite{glenn2015photon,Zhang_PRB2017}, spontaneous quantum synchronization \cite{zhu2015synchronization}, and superradiance \cite{dicke1954coherence,gross1982superradiance}. Some of these also posses signatures of cooperativity. Cascading is however different since the virtual photons are not detected and material resonances are not shifted.

The fifth-order Raman technique uses two pulses. The first creates a vibrational coherence via a Raman process and the second transfers this coherence to another vibrational mode, via another Raman process. The system is finally probed by the transmission of a third pulse after a second variable delay. Fifth-order Raman spectroscopy is a two-dimensional technique that involves two controllable time delays. Cascading occurs when one molecule in the sample serves as a source for inducing the polarization of another molecule. This generates a contribution to the signal that comes as a $\chi^{(3)}\chi^{(3)}$ on top of the desired $\chi^{(5)}$ signal, in that the phase matching given by each lower-order susceptibility in cascading combines to give the same phase-matching condition as the direct $\chi^{(5)}$ process. For example, in one type of six-wave mixing process, light with wavevectors $\textbf{k}_1,\ \textbf{k}_2$ and $\textbf{k}_3$ interact with one molecule via a $\chi^{(3)}$ process to produce a field with $\textbf{k}_v=\textbf{k}_3-\textbf{k}_2+\textbf{k}_1$ and the $\textbf{k}_v$-field together with externally-applied fields $\textbf{k}_4,\ \textbf{k}_5$ interact with another molecule via a second $\chi^{(3)}$ event to produce the signal along the detecting direction $\textbf{k}_s=\textbf{k}_5-\textbf{k}_4+\textbf{k}_v$. 
This cascading signal thus comes in the same direction as the direct signal $\textbf{k}_s=\textbf{k}_5-\textbf{k}_4+\textbf{k}_3-\textbf{k}_2+\textbf{k}_1$. The same argument applies to other choices of signs of $\textbf{k}_j$'s as well as for repeated interactions with fewer pulses.
Cascading obscures the isolation of the desired $\chi^{(5)}$ signal \cite{blank1999,Miller_JCP2008,mehlenbacher2009,golonzka2000,astinov2000diffractive2,kubarych2003} and initial fifth-order Raman experiments in molecular liquid were plagued by cascades \cite{blank1999,golonzka2000,kubarych2003,Miller_JCP2008,tanimura1993two,tominaga1996fifth,tominaga1996temporally,steffen1997time,steffen1997analysis,TokmakoffFleming}. It took several years to recognize the problem of finding out how to eliminate cascading \cite{Wilson_JCP2009,Miller_CPL2000,Mukamel_ARPC2000,Miller_JCP2002,Condon_JPCA2005,Fleming_PRL2002}.
\par
Recent progress in the fabrication of microcavities offers new opportunities for creating dressed matter-photon states known as polaritons. This could lead to entirely new optical properties which significantly modify the chemical landscape \cite{Ebbesen_ACIE2012,Ebbesen_ACIE2016,Yuen_PNAS2011} and molecular properties \cite{kowalewski2016non,kowalewski2016cavity,Zhang_SR2016}. For example, the relaxation dynamics of CO-stretching in W(CO)$_6$ has been modified by strong light-matter coupling, in the pump-probe infrared spectrum \cite{Dunkelberger_NatCommun2016}. It has also been reported that ground-state chemical reactions and photochemical reactivity were significantly slowed down by a cavity \cite{Ebbesen_ACIE2016,Spano_PRL2016}. 

In this article, we demonstrate how cascading processes in fifth-order Raman signals can be manipulated by placing the molecules in an optical microcavity. Intuitively, the coupling of molecules to photons is governed by the mode density of photons, which can be altered in a cavity. Microcavties could thus be used to control the cascading processes. In samples larger than the wavelength of light, the phase-matching condition sets the wavevector of cascading mode and the cavity could be taylored to suppress the density of states at this mode. We explore the relation between cavity geometry and the magnitude of the cascading terms relative to the direct process.  We estimate that the cascading signal in the visible regime can realisticaly be suppressed by $60\%\sim 95\%$ with $\sim$99.5\% suppression a theoretical possibility.

\begin{figure}
 \centering
   \includegraphics[scale=0.25]{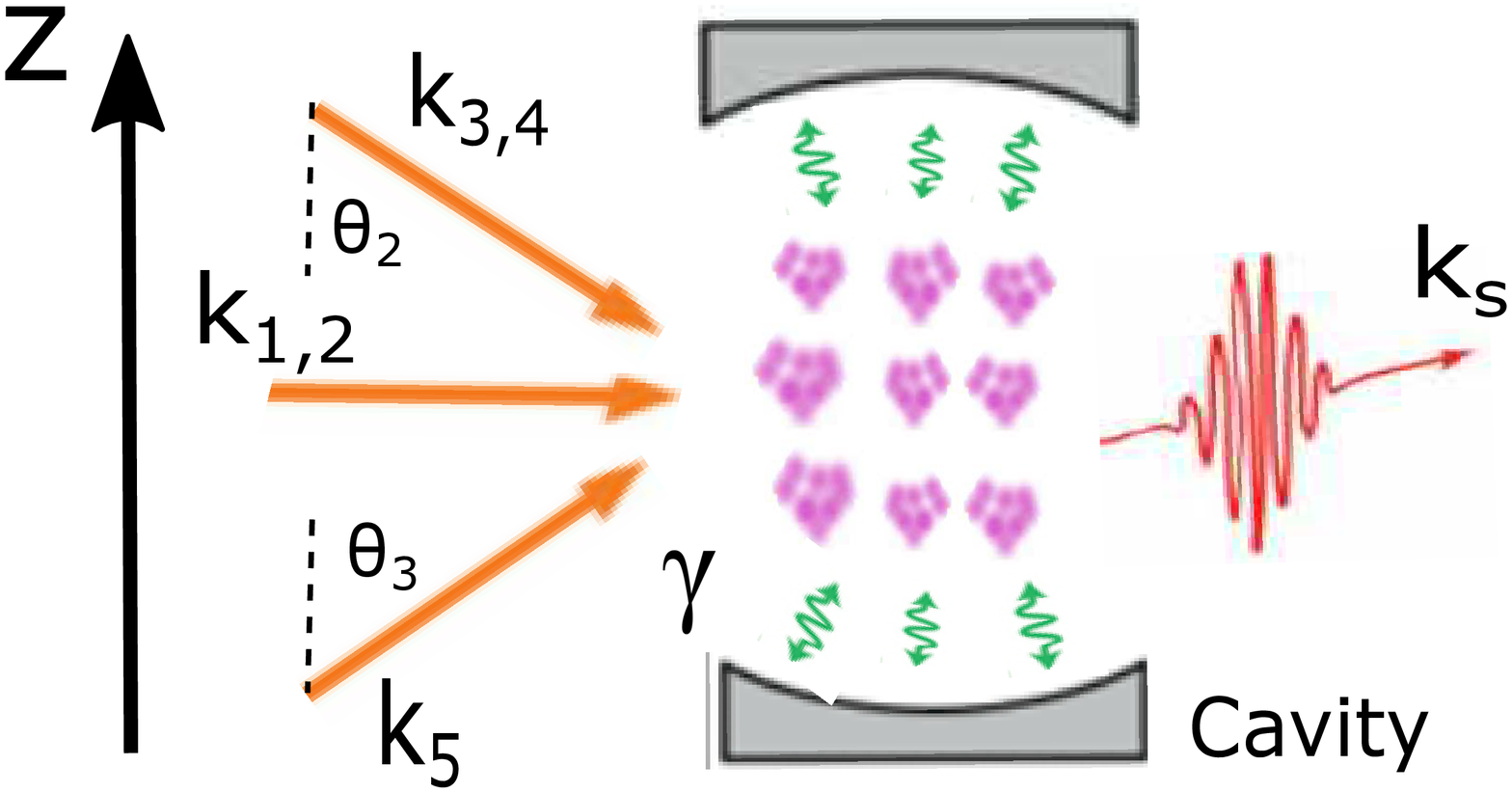}\\[0.5cm]
	 \includegraphics[scale=0.23]{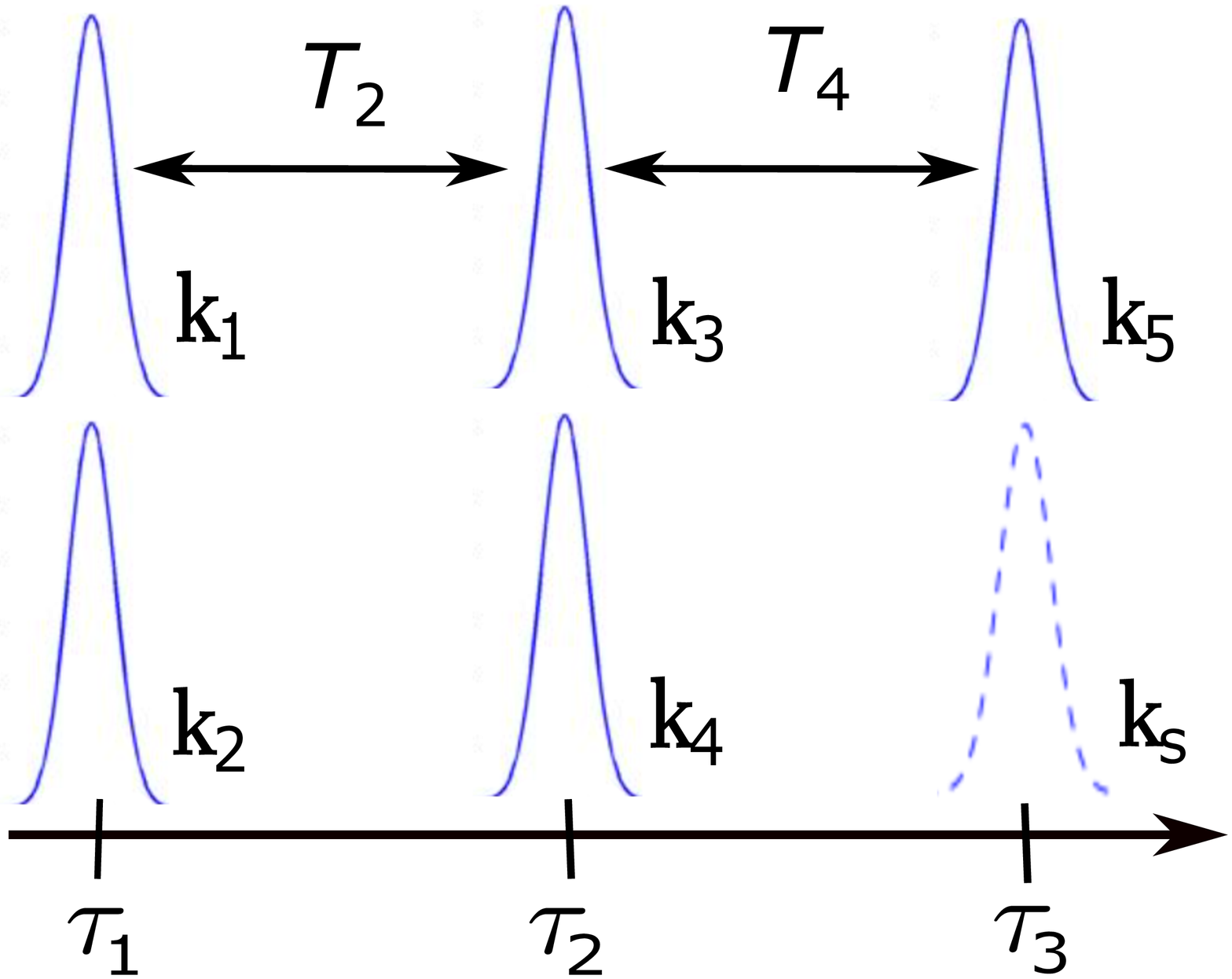}
\caption{(Top) Molecular ensemble interacting with vacuum modes confined in Fabry-Perot microcavity. The photons are confined in the $z$-direction; (Bottom) Pulse sequence of fifth-order Raman spectra.}
\label{sch}
\end{figure}

We consider a homogeneous sample containing $N$ identical molecules in an optical cavity. Each molecule has ground and single-excited electronic levels, accompanied by vibrational manifolds. In a Fabry-Perot cavity, where two mirrors are placed in longitudinal $z$-direction with distance $L$ to access the confinement as shown in Fig.\ \ref{sch}(top), the vacuum modes are quantized with the dispersion relation $\omega_n(\textbf{k}_{\perp})=c\sqrt{|\textbf{k}_{\perp}|^2+\frac{n^2\pi^2}{L^2}}$ with $\textbf{k}_{\perp}$ the wavevector in the transverse $x,y$-direction and $n=1,2,3,\cdots$ denoting the standing wave modes along the $z$-direction. 
The material Hamiltonian reads
\begin{equation}
\begin{split}
H_M = \sum_{a=1}^N\bigg(\sum_{i=1}^{D_g}\varepsilon_g^{(i)}|g_i^{(a)}\rangle\langle g_i^{(a)}|+\sum_{j=1}^{D_e}\varepsilon_e^{(j)}|e_j^{(a)}\rangle\langle e_j^{(a)}|\bigg),
\end{split}
\label{h0}
\end{equation}
where $\vert g_i^{(a)}\rangle$ and $\vert e_i^{(a)}\rangle$ are the $i$th vibrational excitations of the electronic ground and excited states of molecule $a$ respectively while $D_g$ and $D_e$ are the dimensions of the ground and excited vibrational manifolds (the molecules are assumed identical). The photon Hamiltonian is
\begin{equation}
\begin{split}
H_R = \sum_{n=1}^{\infty}\sum_{\textbf{k}_{\perp},\lambda}\hbar\omega_n(\textbf{k}_{\perp})a_{n,\textbf{k}_{\perp}}^{(\lambda),\dagger}a_{n,\textbf{k}_{\perp}}^{(\lambda)},
\end{split}
\label{hr}
\end{equation}
where $a_{n,\textbf{k}_{\perp}}^{(\lambda)}$ is the annihilation operator of the cavity photons and $\lambda$ denotes the photon polarization. The molecule-photon interaction is of the dipolar form $H_{MR} = \sum_{a=1}^N \textbf{P}_a\cdot\textbf{E}(\textbf{r}_a,t)$ with $\textbf{P}_a=\hat{\epsilon}_M^{(a)}(V^++V^-)$ being the dipole moment of molecule $a$ and $\textbf{E}$ is the electric field of the radiation in cavity. $V^-=\sum_{g,e}\mu_{ge}|g^{(a)}\rangle\langle e^{(a)}|,\ V^+\equiv (V^-)^{\dagger}$. With multimode expansion of the electric field,
the molecule-photon interaction can be written as
\begin{equation}
\begin{split}
H_{M\gamma} = \sum_{a=1}^N\sum_{\textbf{k}_{\perp},\lambda}\sum_{n=1}^{\infty} & \left(\hat{\epsilon}_M^{(a)}\cdot\hat{\epsilon}^{(\lambda)}(\textbf{k}_{\perp})\right)\sqrt{\frac{2\pi\omega_n}{\Omega}}\text{sin}\left(\frac{n\pi z_a}{L}\right)\left(V^+ + V^-\right)\\[0.15cm]
& \ \ \times\Big(a_{n,\textbf{k}_{\perp}}^{(\lambda)} e^{i(\textbf{k}_{\perp}\cdot\textbf{r}_a-\omega_n t)} + a_{n,\textbf{k}_{\perp}}^{(\lambda),\dagger}e^{-i(\textbf{k}_{\perp}\cdot\textbf{r}_a-\omega_n t)}\Big),
\end{split}
\label{hint}
\end{equation}
where $\hat{\epsilon}^{(\lambda)}(\textbf{k}_{\perp})$ is the polarization vector of the electric field and $\Omega$ stands for the cavity volume. 

We shall calculate the photon counting signal: $S=\frac{d N}{dt}=\text{Im}[\sum_{a=1}^N\int dt E_s^*(\textbf{r}_a,t)\langle\hat{V}_L(t)\rangle]$ where $V_L$ denotes the transition dipole and is the superoperator acting from the left: $\hat{V}_L\rho\equiv V\rho$. In general, the fifth-order off-resonant Raman signal is induced by five pumping pulses with wave vectors $\textbf{k}_j;\ j=1,2,3,4,5$ and one heterodyne probe with wave vector $\textbf{k}_s$, as shown in Fig.\ \ref{sch}. The signal depends on two time delays $T_2,\ T_4$ as illustrated in Fig.\ \ref{sch}(bottom), making this fifth-order Raman signal a two-dimensional technique. Pulses $\textbf{k}_1$ and $\textbf{k}_2$ are centered at $\bar{\tau}_1$ while the $\textbf{k}_3$-, $\textbf{k}_4$-pulses come at $\bar{\tau}_3$ and $\textbf{k}_5$-, $\textbf{k}_s$-pulses are centered at $\bar{\tau}_5$ ($\bar{\tau}_2\equiv\bar{\tau}_1,\ \bar{\tau}_4\equiv\bar{\tau}_3$). Thus the two delays are $T_2=\bar{\tau}_3-\bar{\tau}_1,\ T_4=\bar{\tau}_5-\bar{\tau}_3$. 
The dipolar field-matter interaction is given by $H_{int}=\sum_{a=1}^N \textbf{P}_a\cdot\textbf{E}(\textbf{r}_a,t)$, with the optical electric field consisting of several pulses
\begin{equation}
\begin{split}
\textbf{E}(\textbf{r},t) = \sum_{j=1}^5 \hat{\epsilon}_j \Big(E_j(t-\bar{\tau}_j)e^{i\left(\textbf{k}_j\cdot\textbf{r}-\omega_j(t-\bar{\tau}_j)\right)} + E_j^*(t-\bar{\tau}_j)e^{-i\left(\textbf{k}_j\cdot\textbf{r}-\omega_j(t-\bar{\tau}_j)\right)}\Big)
\end{split}
\label{pulse}
\end{equation}
and the envelope $E_j(t-\bar{\tau}_j)$ of the $j$-th pulse centered at time $\bar{\tau}_j$ with carrier frequency $\omega_j$ and wavevector $\textbf{k}_j$. The 2D fifth-order Raman signal takes the form of $A\chi^{(5)}+B\chi^{(3)}\chi^{(3)}$ where the first term originates from the direct Raman process since it takes place at the single molecule, and the second term is attributed to cascading. The direct Raman signal is then given by
\begin{equation}
\begin{split}
S_{r}(T_4,T_2) = \text{Im}\bigg[\sum_{a=1}^N\int dt\ (\hat{\epsilon}_s\cdot\hat{\epsilon}_M^{(a)})\text{Tr}\left(E_s(\textbf{r}_a,t)V_L(t) e^{-\frac{i}{\hbar}\int_{-\infty}^t H_{int,-}(\tau)d\tau}\rho(-\infty)\right)\bigg]
\end{split}
\label{Sr}
\end{equation}
where $H_{int,-}(t)=[H_{int}(t),*]$.
Obviously, the direct Raman signal scales as $N$. Substituting Eq.\ (\ref{pulse}) into Eq.\ (\ref{Sr}) and taking the macroscopic limit $\sum_a\rightarrow\frac{N}{\Omega}\int\text{d}^3\textbf{r}$, we finally obtain the fifth-order Raman signal
\begin{equation}
\begin{split}
S_r^{(5)}(T_4,T_2)= & -\frac{4\pi^2 N}{\hbar^5\Omega}\sum_{g,g',g''}\sum_{e,e',e''}\mu_{ge}\mu_{g'e}\mu_{g'e'}\mu_{ge'}\mu_{g''e'}\mu_{g''e''}\\[0.15cm]
& \ \times \mu_{ge''}L\delta^{(2)}(\textbf{k}_s^{\perp}-\textbf{k}_3^{\perp})e^{i\Delta k_z L}\frac{\text{sin}\frac{\Delta k_z L}{2}}{\frac{\Delta k_z L}{2}} \mathcal{M}_{gg'g''}^{ee'e''}(T_4,T_2),
\end{split}
\label{S5st}
\end{equation}
where $\Delta k_z=k_s^z-k_3^z$ is the overall phase mismatching and $L$ is the cavity length . The form of $\mathcal{M}_{gg'g''}^{ee'e''}(T_4,T_2)$ is given in Eq.\ (9) in Supporting Information (SI).

\begin{figure}
 \centering
   \includegraphics[scale=0.27]{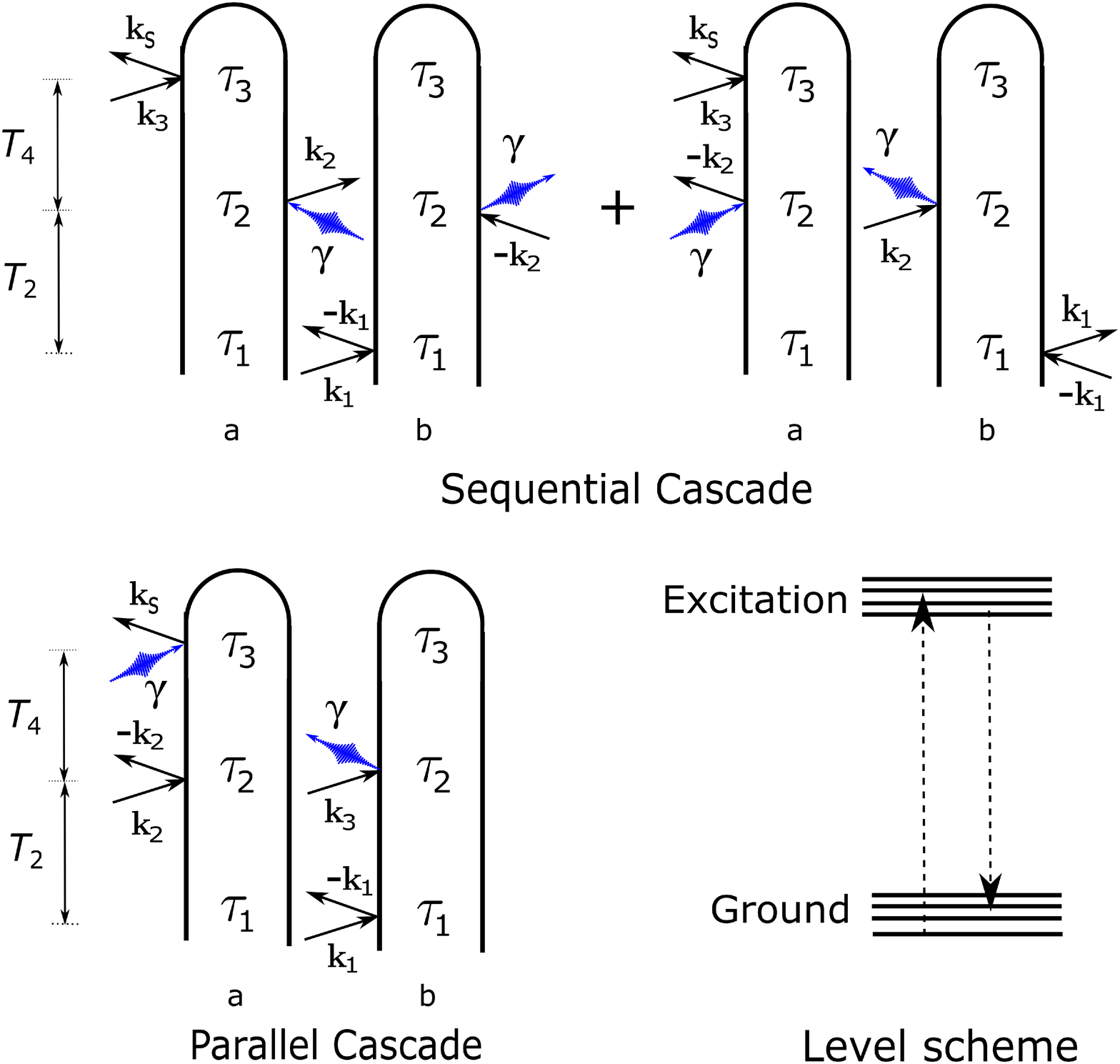}
\caption{Loop diagrams for the sequential and parallel cascades. Black solid and blue wavy arrows stand for the pulses and vacuum modes confined in cavity, respectively; (Bottom right) The vibronic two-level scheme.}
\label{feynman}
\end{figure}

Using Eq.\ (\ref{hint}), the cascading signal calculated to 2nd order in the exciton-photon coupling is
\begin{equation}
\begin{split}
S_c = -\text{Im}\bigg[\sum_{a,b=1}^N\int dt\int_{-\infty}^t d\tau\int_{-\infty}^{\tau}d\tau' E_s^*(\textbf{r}_a,t)\langle V_+(t)V_-(\tau)\rangle_a\langle V_+(\tau')\rangle_b\langle E_{\gamma,+}(\textbf{r}_a,\tau)E_{\gamma,-}(\textbf{r}_b,\tau')\rangle_0\bigg]
\end{split}
\label{Sc}
\end{equation}
As illustrated by the loop diagrams in Fig.\ \ref{feynman}, the fifth-order expansion of Eq.\ (\ref{Sc}) leads to two types of cascading processes. One is the sequential cascading with phase matching $\textbf{k}_s^{\perp}=\textbf{k}_5^{\perp}\pm \textbf{k}_4^{\perp}\mp \textbf{k}_{\perp},\ \textbf{k}_{\perp}=\textbf{k}_3^{\perp}\mp \textbf{k}_2^{\perp}\pm \textbf{k}_1^{\perp}$ and the other is the parallel cascading with phase matching $\textbf{k}_s^{\perp}=\mp \textbf{k}_4^{\perp}\pm \textbf{k}_3^{\perp}+\textbf{k}_{\perp},\ \textbf{k}_{\perp}=\textbf{k}_5^{\perp}\pm \textbf{k}_2^{\perp}\mp \textbf{k}_1^{\perp}$, where $\perp$ denotes the perpendicular $x,y$-direction. The overall fifth-order Raman signal is collected along the following directions: $\textbf{k}_s^{(1)}=\textbf{k}_5+\textbf{k}_4-\textbf{k}_3+\textbf{k}_2-\textbf{k}_1,\ \textbf{k}_s^{(2)}=\textbf{k}_5+\textbf{k}_4-\textbf{k}_3-\textbf{k}_2+\textbf{k}_1,\ \textbf{k}_s^{(3)}=\textbf{k}_5-\textbf{k}_4+\textbf{k}_3+\textbf{k}_2-\textbf{k}_1,\ \textbf{k}_s^{(4)}=\textbf{k}_5-\textbf{k}_4+\textbf{k}_3-\textbf{k}_2+\textbf{k}_1$. The sequential and parallel cascades in the cavity can be obtained by substituting the external pulses Eq.\ (\ref{pulse}) into the cascading signals in Eq.\ (\ref{Sc}) and taking the time-ordering into account. The cascading signals with arbitrary choices of $\textbf{k}_j;\ j=1,2,3,4,5$ of incoming pulses are given in SI. Here we will show the results for $\textbf{k}_2=\textbf{k}_1,\ \textbf{k}_4=\textbf{k}_3$, as done in the experiments for liquid CS$_2$ \cite{blank1999}. The sequential and parallel cascades then take the compact form

\begin{equation}
\begin{split}
& S_{c,\textbf{k}_s=\textbf{k}_3}^{(5),\text{sq}}(T_4,T_2) = -\frac{256\pi^6 N^2}{\hbar^5 \Omega^2}\sum_{g_1,g'_1}\sum_{g_2,g'_2}\sum_{e_1,e'_1}\sum_{e_2,e'_2}\mu_{g_1e_1}\mu_{g'_1e_1}\mu_{g'_1e'_1}\mu_{g_1e'_1}\mu_{g_2e_2}\mu_{g'_2e_2}\mu_{g'_2e'_2}\mu_{g_2e'_2}\\[0.1cm]
& \qquad \times\sum_{m=1}^{\infty}\sum_{\textbf{k}_{\perp}}\frac{\omega_m}{\Omega}\left(\delta^{(2)}(\textbf{k}_2^{\perp}-\textbf{k}_{\perp})\right)^2\frac{m^2\pi^2 L^2}{(\Delta k_{sq}^z L\pm 2m\pi)^2}\left(\frac{\text{sin}\frac{\Delta k_{sq}^z L}{2}}{\frac{\Delta k_{sq}^z L}{2}}\right)^2\times Q_{g_1g'_1e_1e'_1}^m(T_4,T_2)\\[0.35cm]
& S_{c,\textbf{k}_s=\textbf{k}_3}^{(5),\text{pr}} (T_4,T_2) = \frac{512\pi^6 N^2}{\hbar^5 \Omega^2}\sum_{g_1,g'_1}\sum_{g_2,g'_2}\sum_{e_1,e'_1}\sum_{e_2,e'_2}\mu_{g_1e_1}\mu_{g'_1e_1}\mu_{g'_1e'_1}\mu_{g_1e'_1}\mu_{g_2e_2}\mu_{g'_2e_2}\mu_{g'_2e'_2}\mu_{g_2e'_2}\\[0.1cm]
& \qquad \times \sum_{m=1}^{\infty}\sum_{\textbf{k}_{\perp}}\frac{\omega_m}{\Omega}\left(\delta^{(2)}(\textbf{k}_3^{\perp}-\textbf{k}_{\perp})\right)^2\frac{m^2\pi^2 L^2}{(\Delta k_{pr}^z L\pm 2m\pi)^2}\left(\frac{\text{sin}\frac{\Delta k_{pr}^z L}{2}}{\frac{\Delta k_{pr}^z L}{2}}\right)^2\times Y_{g_1g'_1e_1e'_1}^m(T_4,T_2)
\end{split}
\label{Sc5k3}
\end{equation}
where $\Delta k_{sq}^z=k_2^z\mp\frac{m\pi}{L}$ and $\Delta k_{pr}^z=k_3^z\mp\frac{m\pi}{L}$ are the intermediate phase mismatch in the longitudinal direction for sequential and parallel cascades, respectively while $Q_{g_1g'_1e_1e'_1}^m(T_4,T_2)$ and $Y_{g_1g'_1e_1e'_1}^m(T_4,T_2)$ are given in SI to avoid redundancy since the cavity-induced control of cascading signals is dictated by the prefactors in front of $Q$ and $Y$. Since the modes in perpendicular direction are not quantized, the conditions $\textbf{k}_{\perp}\simeq\textbf{k}_2^{\perp}$ and $\textbf{k}_{\perp}\simeq\textbf{k}_3^{\perp}$ can always be satisfied, which leads to the control of cascades by the longitudinal phase mismatch in the prefactor in Eq. (\ref{Sc5k3}). Thus, the photon frequencies are $\omega_m^{sq}=c\sqrt{k_2^2\text{sin}^2\theta_2+\frac{m^2\pi^2}{L^2}}$ and $\omega_m^{pr}=c\sqrt{k_3^2\text{sin}^2\theta_3+\frac{m^2\pi^2}{L^2}}$, where $\theta_2,\ \theta_3$ are the incident angles of $\textbf{k}_2,\ \textbf{k}_3$-pulses with respect to the longitudinal $z$-direction as illustrated in Fig.\ \ref{sch}(top).

The cavity length $L$ must be comparable with the pulse wavelength, namely, $\frac{0.2\pi}{k_2}\lesssim L\lesssim \frac{20\pi}{k_2}$ for sequential and $\frac{0.2\pi}{k_3}\lesssim L\lesssim \frac{20\pi}{k_3}$ for parallel cascades. This is due to the fact that the density of vacuum modes cannot be considerably altered when $L\gg \text{max}\left(\frac{2\pi}{k_2},\frac{2\pi}{k_3}\right)$, which reduces to the free-space case without a avity. We will first discuss the regime $\frac{0.2\pi}{k_2}\lesssim L\lesssim \frac{2\pi}{k_2}$, $\frac{0.2\pi}{k_3}\lesssim L\lesssim \frac{2\pi}{k_3}$ where case the photon frequency is
\begin{equation}
\begin{split}
\omega_m^{sq}\sim ck_2\sqrt{\text{sin}^2\theta_2+\frac{m^2}{4}},\quad \omega_m^{pr}\sim ck_3\sqrt{\text{sin}^2\theta_3+\frac{m^2}{4}}
\end{split}
\label{wm}
\end{equation}
which leads to the estimation of the contributing vacuum modes: $1\le m\lesssim 2$, owing to the resonant condition $\omega_m^{sq},\ \omega_m^{pr}\sim \omega_{eg}\simeq c k_j,\ j=1,2,3$. In the visible regime with wavelength $400\sim 700$nm, the length $L$ of the cavity is $40\text{nm}\lesssim L\lesssim 400\text{nm}$. According to the sinc-function $\frac{\text{sin}^2 x}{x^2}$ in Eq.(\ref{Sc5k3}), the $\gtrsim 50\%$ suppression of the cascades results in $\left | k_2|\text{cos}\theta_2|-\frac{m\pi}{L}\right|\gtrsim\frac{3}{L}$, which gives rise to the range of the angle
\begin{equation}
\begin{split}
|\text{cos}\theta_2|\lesssim\frac{\pi-3}{k_2 L},\quad |\text{cos}\theta_3|\lesssim\frac{\pi-3}{k_3 L}
\end{split}
\label{angle}
\end{equation}
For $L\simeq 100\text{nm}$ and $\lambda_{vis}\simeq 600$nm, the incident angles of $\textbf{k}_2,\ \textbf{k}_3$-pulses can be estimated as $80^o\lesssim\theta_2\lesssim 110^o,\ 80^o\lesssim\theta_3\lesssim 110^o$. This indicates that one can observe the cavity-induced suppression rate of $\gtrsim 50\%$ for cascades in the visible spectrum when orientating the $\textbf{k}_2,\ \textbf{k}_3$ pulses along the direction with $80^o\lesssim\theta_2\lesssim 110^o,\ 80^o\lesssim\theta_3\lesssim 110^o$. Furthermore the maximum suppression rate of $\sim 60\%$ by microcavities is accessible when the signal is collected along the perpendicular direction with $\theta_2,\ \theta_3=90^o$.


We next consider a different scenario where the cavity length $L$ is larger than the wavelength of the pulses, specifically, $L\sim \frac{2p\pi}{k_i},\ i=2,3$ and $1\lesssim p\lesssim 10$. In this case, the frequencies of the vacuum photons for squential and parallel cascades are
\begin{equation}
\begin{split}
\omega_m^{sq}\simeq ck_2\sqrt{\text{sin}^2\theta_2+\frac{m^2}{4p^2}},\ \omega_m^{pr}\simeq ck_3\sqrt{\text{sin}^2\theta_3+\frac{m^2}{4p^2}}
\end{split}
\label{wml}
\end{equation}
which gives rise to the estimation of the contributing vacuum modes: $1\le m\lesssim 2p$, owing to the resonant condition $\omega_m^{sq},\ \omega_m^{pr}\sim \omega_{eg}\simeq c k_j,\ j=1,2,3$. Based on the property of the sinc-function in the prefactor in Eq.(\ref{Sc5k3}) the suppression of cascades with the ratio $\gtrsim 95\%$ demands $\left| k_i L|\text{cos}\theta_i|-m\pi\right|\gtrsim 5$ which leads to $|\text{cos}\theta_i|\gtrsim\frac{2p\pi+5}{k_i L}$. By setting $L\simeq \frac{2\pi}{k_i}(p+1)$ we obtain the estimated range for angle
\begin{equation}
\begin{split}
|\text{cos}\theta_i|\gtrsim\frac{p+\frac{5}{2\pi}}{p+1};\ i=2,3
\end{split}
\label{anglel}
\end{equation}

For the situation when $L\simeq \frac{4\pi}{k_i}$ giving $p\simeq 1$ (i.e., $\lambda_{vis}\simeq 500\text{nm}$ in visible spectrum, $L\simeq 1\mu\text{m}$), only $m=1,2$ contribute to the summation over $m$ in Eq.(\ref{Sc5k3}), which results in the observation of $\gtrsim 95\%$ suppression of cascades when the signal is collected along the direction $\theta_3\lesssim 26^o$ with the orientation $\theta_2\lesssim 26^o$ of the $\textbf{k}_2$-signal. It is worth noticing that a $\sim 99.5\%$ suppression of cascades can be achieved when the $\textbf{k}_2$- and $\textbf{k}_3$-pulses are orientated along the cavity axis ($z$-direction here) and $k_i L=2(p+1)\pi$, due to the fact that the upper bound of the dimensionless prefactor of $m=1$ term in the summation in Eq.(\ref{Sc5k3}) reads $\frac{\pi^2}{(4\pi-\pi)^2}\times\frac{\text{sin}^2\frac{\pi}{2}}{(2\pi-\frac{\pi}{2})^2}\simeq 0.005$.

In conclusion we demonstrated that the cascading processes can be considerably suppressed by controlling the size of microcavity and selecting the direction of the incoming pulses. Our suppression scheme operates by altering the electromagnetic density of states from its free-space value, in particular in the vicinity of third-order linear combinations of incoming wavevectors (see discussion after Eq.\ (\ref{Sc})). A numerical estimation of the cavity geometry for visible light shows that the cascading signal can be greatly suppressed, in principle up to $\gtrsim 99.5\%$. Previously, the contamination of this intermediate process was shown to be reduced by the design of polarization configurations, i.e., Dutch Cross, which could achieve a suppression of four orders of magnitude \cite{Jansen_JCP2001,Miller_CPL2003}. These existing designs could be combined with a cavity-suppression scheme to overcome cascading in dense samples. Our scheme also suggests further avenues for manipulation of the cascading processes by, e.g., using multiple, resonantly-coupled cavities rather than a single cavity or otherwise spatially modulating the cavity structure.  
Our results may offer a new route to manipulating the cascading processes, which plays an important role in multidimensional spectroscopy.

\begin{acknowledgement}
We gratefully thank the support of the National Science Foundation (grant CHE-1361516) and of the Chemical Sciences, Geosciences, and Biosciences division, Office of Basic Energy Sciences, Office of Science, U.S. Department of Energy through award No. DE-FG02-04ER15571. Support for K.B. was provided by DOE.
\end{acknowledgement}


\begin{suppinfo}
The Supporting Information is available free of charge on the ACS Publication website at DOI:
\end{suppinfo}


\begin{thebibliography}{}
\bibitem{blank1999}
Blank, D.\ A.; Kaufman, L.\ J.; Fleming, G.\ R. Fifth-order Two-dimensional Raman Spectra of CS$_2$ Are Dominated by Third-order Cascades. {\it J. Chem. Phys.} \textbf{1999}, 111, 3105-3114
\bibitem{mehlenbacher2009}
Mehlenbacher, R.\ D.; Lyons, B.; Wilson, K.\ C.; Du, Y.; McCamant, D.\ W. Theoretical Analysis of Anharmonic Coupling and Cascading Raman Signals Observed with Femtosecond Stimulated Raman Spectroscopy. {\it . Chem.
Phys.} \textbf{2009}, 131, 244512-244532
\bibitem{Miller_JCP2008}
Li, Y.; Huang, L.; Hasegawa, M.; Tanimura, Y.; Dwayne Miller, R.\ J. Two-Dimensional Fifth-Order Raman Spectroscopy of Liquid Formamide: Experiment and Theory. {\it J. Chem. Phys.} \textbf{2008}, 128, 234507-234521
\bibitem{golonzka2000}
Golonzka, O.; Demird$\ddot{\text{o}}$ven, N.; Khalil, M.; Tokmakoff, A. Separation of Cascaded and Direct Fifth-order Raman Signals using Phase-sensitive Intrinsic Heterodyne Detection. {\it J. Chem. Phys.} \textbf{2000}, 113, 9893-9896
\bibitem{zhao2011}
Zhao, B.; Sun, Z.\ G.; Lee, Soo-Y. Quantum Theory of Time-resolved Femtosecond Stimulated Raman Spectroscopy: Direct versus Cascade Processes and Application to CDCl$_3$. {\it J. Chem. Phys.} \textbf{2011}, 134, 024307-024318
\bibitem{frostig2015single}
Frostig, H.; Bayer, T.; Dudovich, N.; Eldar, Y.\ C.; Silberberg, Y. Single-beam Spectrally Controlled Two-dimensional Raman Spectroscopy. {\it Nat. Photonics} \textbf{2015}, 9, 339-343
\bibitem{gelin2013simple}
Gelin, M.\ F.; Domcke, W. Simple Recipes for Separating Excited-state Absorption and Cascading Signals by Polarization-sensitive Measurements. {\it J. Phys. Chem. A} \textbf{2013}, 117, 11509-11513
\bibitem{cascadingQED}
Bennett, K.; Chernyak, V.\ Y.; Mukamel, S. Discriminating Cascading Processes in Nonlinear Optics: A QED Analysis Based on Their Molecular and Geometric Origin. {\it Phys. Rev. A} \textbf{2017}, 95, 033840-033852
\bibitem{bennett2014cascading}
Bennett, K.; Mukamel, S. Cascading and Local-field Effects in Nonlinear Optics Revisited: A Quantum-field Picture Based on Exchange of Photons. {\it J. Chem. Phys.} \textbf{2014}, 140, 044313-044323
\bibitem{mukbook}
Mukamel, S. {\it Principles of Nonlinear Optical Spectroscopy}; Oxford University Press: New York, U.S.A., 1995
\bibitem{lozovoy2001cascaded}
Lozovoy, V.\ V.; Pastirk, I.; Comstock, M.\ G.; Dantus, M. Cascaded Free-induction Decay Four-wave Mixing. {\it Chem. Phys.} \textbf{2001}, 266, 205-212
\bibitem{cundiff2002time}
Cundiff, S.\ T. Time Domain Observation of the Lorentz-Local Field. {\it Laser Phys.} \textbf{2002}, 12, 1073-1078
\bibitem{thirunamachandran1980intermolecular}
Thirunamachandran, T. Intermolecular Interactions in the Presence of An Intense Radiation Field. {\it Mol. Phys.} \textbf{1980}, 40, 393-399
\bibitem{salam}
Salam, A. {\it Molecular Quantum Electrodynamics: Long-Range Intermolecular Interactions}; Wiley: New York, U.S.A., 2010
\bibitem{scully2009collective}
Scully, M.\ O. Collective Lamb Shift in Single Photon Dicke Superradiance. {\it Phys. Rev. Lett.} \textbf{2009}, 102, 143601-143604
\bibitem{glenn2015photon}
Glenn, R.; Bennett, K.; Dorfman, K.\ E.; Mukamel, S. Photon-exchange Induces Optical Nonlinearities in Harmonic Systems. {\it J. Phys. B: Atomic, Molecular and Optical Physics} \textbf{2015}, 48, 065401-065420
\bibitem{Zhang_PRB2017}
Zhang, Z.\ D.; Fu, H.\ C.; Wang, J. Nonequilibrium-induced Enhancement of Dynamical Quantum Coherence and Entanglement of Spin Arrays. {\it Phys. Rev. B} \textbf{2017}, 95, 144306-144314
\bibitem{zhu2015synchronization}
Zhu, B.; Schachenmayer, J.; Xu, M.; Herrera, F.; Restrepo, J.\ G.; Holland, M.\ J.; Rey, A.\ M. Synchronization of Interacting Quantum Dipoles. {\it New J. Phys.} \textbf{2015}, 17, 083063-083077
\bibitem{dicke1954coherence}
Dicke, R.\ H.; Coherence in Spontaneous Radiation Processes. {\it Phys. Rev.} \textbf{1954}, 93, 99-110
\bibitem{gross1982superradiance}
Gross, M.; Haroche, S. Superradiance: An Essay on The Theory of Collective Spontaneous Emission. {\it Phys. Rep.} \textbf{1982}, 93, 301-396
\bibitem{astinov2000diffractive2}
Astinov, V.; Kubarych, K.\ J.; Milne, C.\ J.; Miller, R.\ J.\ D. Diffractive Optics Implementation of Six-wave Mixing. {\it Opt. Lett.} \textbf{2000}, 25, 853-855
\bibitem{kubarych2003}
Kubarych, K.\ J.; Milne, C.\ J.; Miller, R.\ J.\ D. Fifth-order Two-dimensional Raman Spectroscopy: A New Direct Probe of The Liquid State. {\it Int. Rev. Phys. Chem.} \textbf{2003}, 22, 497-532
\bibitem{tanimura1993two}
Tanimura, Y.; Mukamel, S. Two-dimensional Femtosecond Vibrational Spectroscopy of Liquids. {\it J. Chem. Phys.} \textbf{1993}, 99, 9496-9511
\bibitem{tominaga1996fifth}
Tominaga, K.; Yoshihara, K. Fifth-order Nonlinear Spectroscopy on The Low-frequency Modes of Liquid CS$_2$. {\it J. Chem. Phys.} \textbf{1996}, 104, 4419-4426
\bibitem{tominaga1996temporally}
Tominaga, K.; Yoshihara, K. Temporally Two-dimensional Femtosecond Spectroscopy of Binary Mixture of CS$_2$. {\it J. Chem. Phys.} \textbf{1996}, 104, 1159-1162
\bibitem{steffen1997time}
Steffen, T.; Duppen, K. Time Resolved Four-and Six-wave Mixing in Liquids II: Experiments. {\it J. Chem. Phys.} \textbf{1997}, 106, 3854-3864
\bibitem{steffen1997analysis}
Steffen, T.; Duppen, K. Analysis of Nonlinear Optical Contributions to Temporally Two-dimensional Raman Scattering. {\it Chem. Phys. Lett.} \textbf{1997}, 273, 47-54
\bibitem{TokmakoffFleming}
Tokmakoff, A.; Fleming, G.\ R. Two-dimensional Raman Spectroscopy of The Intermolecular Modes of Liquid CS$_2$. {\it J. Chem. Phys.} \textbf{1997}, 106, 2569-2582
\bibitem{Wilson_JCP2009}
Wilson, K.\ C.; Lyons, B.; Mehlenbacher,R.; Sabatini,R.; McCamant,D.\ W. Two-dimensional Femtosecond Stimulated Raman Spectroscopy: Observation of Cascading Raman Signals in Acetonitrile. {\it J. Chem. Phys.} \textbf{2009}, 131, 214502-214516
\bibitem{Miller_CPL2000}
Astinov, V.; Kubarych, K.\ J.; Milne, C.\ J.; Miller, R.\ J.\ D. Diffractive Optics Based Two-color Six-wave Mixing: Phase Contrast Heterodyne Detection of The Fifth Order Raman Response of Liquids. {\it Chem. Phys. Lett.} \textbf{2000}, 327, 334-342
\bibitem{Mukamel_ARPC2000}
Mukamel, S. Multidimensional Femtosecond Correlation Spectroscopies of Electronic and Vibrational Excitations. {\it Annu. Rev. Phys. Chem.} \textbf{2000}, 51, 691-731
\bibitem{Miller_JCP2002}
Kubarych, K.\ J.; Milne, C.\ J.; Lin,S.; Astinov,V.; Miller,R.\ J.\ D. Diffractive Optics-based Six-wave Mixing: Heterodyne Detection of The Full $\chi^{(5)}$ Tensor of Liquid CS$_2$. {\it J. Chem. Phys.} \textbf{2002}, 116, 2016-2042
\bibitem{Condon_JPCA2005}
Condon, N.\ J.; Wright, J.\ C. Doubly Vibrationally Enhanced Four-Wave Mixing in Crotononitrile. {\it J. Phys. Chem. A} \textbf{2005}, 109, 721-729
\bibitem{Fleming_PRL2002}
Kaufman,L.\ J.; Heo, J.; Ziegler, L.\ D.; Fleming, G.\ R. Heterodyne-Detected Fifth-Order Nonresonant Raman Scattering from Room Temperature CS$_2$. {\it Phys. Rev. Lett.} \textbf{2002}, 88, 207402-207405
\bibitem{Ebbesen_ACIE2012}
Hutchison, J.\ A.; Schwartz, T.; Genet, C.; Devaux, E.; Ebbesen, T.\ W. Modifying Chemical Landscapes by Coupling to Vacuum Fields. {\it Angew. Chem. Int. Ed.} \textbf{2012}, 124, 1624-1628
\bibitem{Ebbesen_ACIE2016}
Thomas, A.; George, J.; Shalabney, A.; Dryzhakov, M.; Varma, S.\ J.; Moran, J.; Chervy, T.; Zhong, X.; Devaux, E.; Genet, C.; Hutchison, J.\ A.; Ebbesen, T.\ W. Ground-state Chemical Reactivity under Vibrational Coupling to the Vacuum Electromagnetic Field. {\it Angew. Chem. Int. Ed.} \textbf{2016}, 55, 11462-11466
\bibitem{Yuen_PNAS2011}
Yuen-Zhou, J.; Krich, J.\ J.; Mohseni, M.; Aspuru-Guzik, A. Quantum State and Process Tomography of Energy Transfer Systems via Ultrafast Spectroscopy. {\it Proc. Nat. Acad. Sci. U.S.A.} \textbf{2011}, 108, 17615-17620
\bibitem{kowalewski2016non}
Kowalewski, M.; Bennett, K.; Mukamel, S. Non-adiabatic Dynamics of Molecules in Optical Cavities. {J. Chem. Phys.} \textbf{2016}, 144, 054309-054316
\bibitem{kowalewski2016cavity}
Kowalewski, M.; Bennett, K.; Mukamel, S. Cavity Femtochemistry; Manipulating Nonadiabatic Dynamics at Avoided Crossings. {\it J. Phys. Chem. Lett.} \textbf{2016}, 7, 2050-2054
\bibitem{Zhang_SR2016}
Zhang, Z.\ D.; Wang, J. Origin of Long-Lived Quantum Coherence and Excitation Dynamics in Pigment-Protein Complexes. {\it Sci. Rep.} \textbf{2016}, 6, 37629-37637
\bibitem{Dunkelberger_NatCommun2016}
Dunkelberger,A.\ D.; Spann, B.\ T.; Fears, K.\ P.; Simpkins, B.\ S.; Owrutsky, J.\ C. Modified Relaxation Dynamics and Coherent Energy Exchange in Coupled Vibration-cavity Polaritons. {\it Nat. Commun.} \textbf{2016}, 7, 13504-13513
\bibitem{Spano_PRL2016}
Herrera, F.; Spano, F.\ C. Cavity-Controlled Chemistry in Molecular Ensembles. {\it Phys. Rev. Lett.} \textbf{2016}, 116, 238301-238305
\bibitem{Jansen_JCP2001}
Jansen, T.\ I.\ C.; Snijders, J.\ G.; Duppen, K. Interaction Induced Effects in the Nonlinear Raman Response of Liquid CS$_2$: A Finite Field Nonequilibrium Molecular Dynamics Approach. {\it J. Chem. Phys.} \textbf{2001}, 114, 10910-10921
\bibitem{Miller_CPL2003}
Kubarych, K.\ J.; Milne, C.\ J.; Miller, R.\ J.\ D. Heterodyne Detected Fifth-order Raman Response of Liquid CS$_2$: Dutch Cross Polarization. {\it Chem. Phys. Lett.} \textbf{2003}, 369, 635-642
\end{thebibliography}


\end{document}